\documentclass[10pt,conference,twocolumn]{IEEEtran}

\usepackage[noadjust]{cite}
\usepackage{amssymb, amsmath}

\renewcommand{\QED}{\QEDopen}

\newtheorem{theorem}{Theorem}
\newtheorem{lemma}[theorem]{Lemma}
\newtheorem{corollary}[theorem]{Corollary}
\newtheorem{result}[theorem]{Result}
\newtheorem{definition}[theorem]{Definition}

\newcommand{\beqn}{\begin{equation}}
\newcommand{\eeqn}{\end{equation}}
\newcommand{\beq}{\begin{equation*}}
\newcommand{\eeq}{\end{equation*}}
\newcommand{\Z}{\mathbb Z}
\newcommand{\C}{{\cal C}}
\newcommand{\A}{{\cal A}}
\renewcommand{\L}{{\cal L}}
\newcommand{\R}{{\cal R}}

\newcommand{\RM}{{\rm RM}}
\newcommand{\ZRM}{{\rm ZRM}}
\newcommand{\PMEPR}{\mbox{PMEPR}}

\newcommand{\wt}{{\rm wt}}
\newcommand{\dist}{{\rm d}}

\renewcommand{\emptyset}{\varnothing}

\renewcommand{\th}{{\rm th}}

\newcommand{\bia}{{\boldsymbol{a}}}
\newcommand{\bib}{{\boldsymbol{b}}}
\newcommand{\bif}{{\boldsymbol{f}}}
\newcommand{\bix}{{\boldsymbol{x}}}
\newcommand{\bic}{{\boldsymbol{c}}}
\newcommand{\bid}{{\boldsymbol{d}}}
\newcommand{\biA}{{\boldsymbol{A}}}
\newcommand{\biB}{{\boldsymbol{B}}}
\newcommand{\biF}{{\boldsymbol{F}}}
\newcommand{\biG}{{\boldsymbol{G}}}
\newcommand{\bnull}{{\boldsymbol{0}}}

\begin{document}

\title{New Codes for OFDM with Low PMEPR}%
\author{Kai-Uwe Schmidt and Adolf Finger
\\
Communications Laboratory\\ 
Dresden University of Technology\\
01062 Dresden, Germany\\
Email: schmidtk@ifn.et.tu-dresden.de}

\maketitle

\thispagestyle{empty}

\begin{abstract}
In this paper new codes for orthogonal frequency-division multiplexing (OFDM) with tightly controlled peak-to-mean envelope power ratio (PMEPR) are proposed. We identify a new family of sequences occuring in complementary sets and show that such sequences form subsets of a new generalization of the Reed--Muller codes. Contrarily to previous constructions we present a compact description of such codes, which makes them suitable even for larger block lengths. We also show that some previous constructions just occur as special cases in our construction.
\end{abstract}


\section{Introduction}
\PARstart{L}{et} us consider an $n$-subcarrier orthogonal frequency-division multiplexing (OFDM) system. The signal
\beq
s_\biA(t)=\sum_{i=0}^{n-1}A_i e^{2\pi\sqrt{-1} (f_c+if_s) t}\quad (0\le t<T)
\eeq
is called the complex envelope of the transmitted signal. Here $T$ denotes the symbol duration, $f_s$ is the subcarrier spacing, and $f_c$ is the radio carrier frequency. 
In an ideal situation it is commonly assumed that $f_s=1/T$. 
The vector $\biA=(A_0\,A_1\dots A_{n-1})$ is called the modulating codeword of the OFDM symbol. Let us assume that each subcarrier is modulated with a $q$-ary phase-shift-keying (PSK) constellation. 
Our concern is the envelope power of the transmitted signal $P_\biA(t)=|s_\biA(t)|^2$. An important characteristic of an OFDM signal is the {\em peak-to-mean envelope power ratio} (PMEPR), which is for PSK-modulated subcarriers defined as
\beq
\PMEPR(\biA)=\frac{1}{n}\sup_{0\le t<T}P_\biA(t).
\eeq
For uncoded transmission the PMEPR is typically much higher than $1$ and can grow up to as much as $n$. The high PMEPR of uncoded OFDM signals can be considered as the major drawback of the OFDM technique. Due to the high signal dynamics, the power amplifier should have a large linear range causing inefficient operation. On the other hand, a nonlinear power amplifier may result in severe signal distortion, such as interferences between the subcarriers and out-of-band radiation, where the latter issue is subject to strong regulations.
\par
There exists a number of approaches to alleviate the problem of high PMEPR. A promising one remains the use of coding across the subcarriers \cite{ Jones1996}. The employed code should comprise only those codewords having low PMEPR, and in addition, it should provide a certain level of error protection. Let $\C$ denote such a code and define the PMEPR of the code $\C$
\beq
\PMEPR(\C):=\max_{\biA\in\C}\PMEPR(\biA).
\eeq
So, for a given $n$, we aim to find codes with low PMEPR, good error protection, and high rate. In \cite{Davis1999} good codes with PMEPR at most $2$ were constructed for small $n$ by establishing a link between Golay complementary pairs \cite{Golay1961} and certain second-order cosets of a generalized first-order Reed--Muller code. This technique was extended and generalized in \cite{Paterson2000a} by including sequences lying in complementary sets \cite{Tseng1972}. However the codes are still unions of quadratic cosets of a generalized first-order Reed--Muller code and are a bit unwieldy, which makes them only suitable for small $n$. Recently, in \cite{Schmidt2005}, progress has been made in constructing sequences lying in complementary sets, which are not necessarily of quadratic order. These sequences in connection with new generalizations of the classical Reed--Muller code will be used in this paper to build powerful codes with bounded PMEPR and good error protection properties.
\par
The remainder of this paper is organized as follows. In the next section we merely adopt some useful notation. In Section \ref{sec:sequences} we present a new family of complementary sequences. A new generalization of the Reed--Muller codes is introduced in Section \ref{sec:RM-codes}. In Section \ref{sec:ofdm-codes} we present our code constructions. Section \ref{sec:conclusion} concludes the paper.


\section{Notation and Preliminaries}
\label{sec:notation}
Let $\biA=(A_0\,A_1\,\cdots\,A_{n-1})$ and $\biB=(B_0\,B_1\,\cdots\,B_{n-1})$ be two complex-valued vectors. Then the aperiodic cross-correlation of $\biA$ and $\biB$ at a displacement $\ell$ is given by
\beq
C(\biA,\biB)(\ell):=\left\{
\begin{array}{ll}
\sum\limits_{i=0}^{n-\ell-1}A_{i+\ell}B^*_i&0\le \ell<n\\
\sum\limits_{i=0}^{n+\ell-1}A_iB^*_{i-\ell}&-n< \ell<0\\
0&\mbox{otherwise}
\end{array}\right.,
\eeq
where $()^*$ denotes complex conjugation. 
The aperiodic auto-correlation of $\biA$ at a displacement $\ell$ is then conveniently written as 
\beq
A(\biA)(\ell):=C(\biA,\biA)(\ell).
\eeq
\par
A generalized Boolean function $f$ is defined as a mapping $f\,:\,\Z_2^m\rightarrow\Z_q$. Such a function can be uniquely written in its algebraic normal form, i.e. $f$ is a sum of the $2^m$ weighted monomials
\beq
f=f(x_0,x_1,\dots,x_{m-1})=\sum_{i=0}^{2^m-1}c_i\,\prod_{\alpha=0}^{m-1} x_\alpha^{i_\alpha},
\eeq
where $c_0,\dots,c_{2^m-1}\in\Z_q$ and $(i_0\,i_1\dots i_{m-1})$ is the binary expansion of the integer $i$, such that $i=\sum_{j=0}^{m-1}i_j2^j$. The order of the $i\th$ monomial is defined as $\sum_{j=0}^{m-1}i_j$, and the order of a generalized Boolean function is equal to the highest order of the monomials with a nonzero coefficient in the algebraic normal form of $f$. 
\par
A generalized Boolean function may be equally represented by vectors of length $2^m$. We shall define the vector $\bif=(f_0\,f_1\cdots f_{2^m-1})$ and the vector $\biF=\xi^\bif=(\xi^{f_0}\,\xi^{f_1}\cdots \xi^{f_{2^m-1}})$ as the $\Z_q$-valued vector and the polyphase vector associated with $f$, respectively. Here $\xi=\exp(2\pi\sqrt{-1}/q)$ is a primitive $q\th$ root of unity, and $f_i=f(i_0,i_1,\cdots,i_{m-1})$, where $(i_0\,i_1\cdots i_{m-1})$ is the binary expansion of the integer $i$. Throughout this paper $q$ is assumed to be even.
\par
We shall now define the restriction of polyphase vectors of length $2^m$ and their corresponding generalized Boolean functions. This technique was  introduced in \cite{Paterson2000a} and it will be useful to prove the results in this paper. Let $f: \Z_2^m\rightarrow\Z_q$ be a generalized Boolean function in the variables $x_0,x_1,\cdots,x_{m-1}$, and let $\biF$ be its associated polyphase vector. Suppose $0\le j_0<j_1<\cdots<j_{k-1}<m$ is a list of $k$ indices and write $\bix=(x_{j_0}\,x_{j_1}\,\cdots\,x_{j_{k-1}})$. Let $\bid=(d_0\,d_1\,\cdots\,d_{k-1})$ be an arbitrary binary vector of length $k$, and let  $(i_0\,i_1\,\cdots\,i_{m-1})$ be the binary expansion of the integer $0\le i<2^m$. Then the restricted vector $\biF|_{\bix=\bid}$ is a vector of length $2^m$ with its elements $(\biF|_{\bix=\bid})_i$ ($i=0,1,\cdots,2^m-1$) being defined as
\beq
(\biF|_{\bix=\bid})_i:=\left\{
\begin{array}{ll}
F_i&\mbox{if}\;(i_{j_0}\,i_{j_1}\cdots i_{j_{k-1}})=(d_0\,d_1\cdots d_{k-1})\\
0&\mbox{if}\;(i_{j_0}\,i_{j_1}\cdots i_{j_{k-1}})\ne(d_0\,d_1\cdots d_{k-1})
\end{array}\right.\!\!.
\eeq
For the case $k=0$ we fix $\biF|_{\bix=\bid}=\biF$.\label{def:restriction}
\par
A vector that is restricted in $k$ variables comprises $2^m-2^{m-k}$ zero entries and $2^{m-k}$ nonzero entries. Those nonzero entries are determined by a function, which we shall denote as $f|_{\bix=\bid}$. This function is a Boolean function in $m-k$ variables and is obtained by replacing the variables $x_{j_\alpha}$ by $d_\alpha$ for all $0\le\alpha<k$ in the original function $f$. The restricted vector $\biF|_{\bix=\bid}$ is then found by associating a polyphase vector of length $2^{m-k}$ with $f|_{\bix=\bid}$ and inserting $2^m-2^{m-k}$ zeros at the corresponding positions. Similarly to a disjunctive normal form of a Boolean function \cite{MacWilliams1977}, the original function $f$ can be reconstructed from the functions $f|_{\bix=\bid}$ by
\beq
f=\sum_\bid f|_{\bix=\bid}\prod_{i=0}^{k-1}x_{j_i}^{d_i}(1-x_{j_i})^{(1-d_i)}.
\eeq
\par
\begin{lemma}\cite{Paterson2000a},\cite{Stinchcombe2000}
\label{lem:CCF-Expanding}
Let $f:\Z_2^m\rightarrow \Z_q$ be a generalized Boolean function in the variables $x_0,x_1,\cdots ,x_{m-1}$, and let $\biF$ be its associated polyphase vector. Let $0\le j_0<j_1<\cdots<j_{k-1}<m$ be a list of $k$ indices and write $\bix=(x_{j_0}\,x_{j_1}\,\cdots\,x_{j_{k-1}})$. Suppose $\bid,\bid_1,\bid_2$, are binary vectors of length $k$. Then we have
\beq
A(\biF)(\ell)=\sum\limits_{\bid}A(\biF|_{\bix=\bid})(\ell)\!+\!\!\!
\sum\limits_{\bid_1\ne\bid_2}C(\biF|_{\bix=\bid_1},\biF|_{\bix=\bid_2})(\ell).
\eeq
\end{lemma}


\section{Complementary Sequences}
\label{sec:sequences}
\begin{definition}\cite{Tseng1972},\cite{Paterson2000a}
A set of $N$ sequences is called a complementary set of size $N$ if the aperiodic auto-correlations of its members sum up to zero except for the zero displacement. If $N=2$, the two sequences are commonly termed a Golay complementary pair \cite{Golay1961}.
\end{definition}
\begin{result}\cite{Paterson2000a}
\label{res:sets}
The PMEPR of a sequence lying in a complementary set of size $N$ is at most $N$.
\end{result}
\par
The above result motivates the construction of sequences lying in complementary sets of small size and use them as codewords in OFDM. For the sake of efficiency, we require an explicit construction method that generates many of them.
\par
A main result of \cite{Paterson2000a}, generalizing the work in \cite{Davis1999}, is the following theorem, which describes the construction of complementary pairs.
\begin{theorem}\cite{Paterson2000a}
\label{thm:pairs}
Let $ J=\{j_0,j_1,\cdots,j_{k-1}\}$ and $I=\{i_0,i_1,\cdots,i_{m-k-1}\}$ be two sets of indices, such that $I\cap J=\emptyset$ and $I\cup J=\{0,1,\cdots,m-1\}$.
Write $\bix=(x_{j_0}\,x_{j_1}\, \cdots\, x_{j_{k-1}})$, and let $\bid$ be an arbitrary binary vector of length $k$. Let $f: \Z_2^m\rightarrow\Z_q$ be a generalized Boolean function in $m$ variables $x_{0},x_{1},\cdots,x_{{m-1}}$, such that $f|_{\bix=\bid}$ is of the form
\beqn
\frac{q}{2}\sum_{\alpha=0}^{m-k-2}x_{\pi(\alpha)}x_{\pi(\alpha+1)}+\sum_{\alpha=0}^{m-k-1}c_\alpha x_{i_\alpha}+c,
\label{eqn:pairs}
\eeqn
where $c_0,\cdots,c_{m-k-1},c\in\Z_q$ and $\pi$ is a permutation of the indices $\{i_0,i_1,\cdots,i_{m-k-1}\}$. Let $\biF$ and $\biF'$ be the polyphase vectors associated with the functions $f$ and $f+(q/2)x_a+c'$, respectively. Here $a$ is either $\pi(i_0)$ or $\pi(i_{m-k-1})$ and $c'\in\Z_q$. Then $\biF|_{\bix=\bid}$ and $\biF'|_{\bix=\bid}$ comprise a complementary pair.
\end{theorem}
\par
{\itshape Remark:} 
In particular, if $k=0$, the above theorem identifies  $(m!/2)q^{m+1}$ polyphase sequences lying in complementary pairs. For $q$ being a power of $2$ these are exactly those constructed in \cite{Davis1999}.
\par
The following theorem was recently obtained in \cite{Schmidt2005} and describes the construction of complementary sets. It will be crucial for our new code constructions in Section \ref{sec:ofdm-codes}.
It generalizes \cite[Theorem 12]{Paterson2000a} from sequences associated with quadratic generalized Boolean functions to sequences associated with generalized Boolean functions of arbitrary order.
\par
\begin{theorem}
\label{thm:sets}
Define two index sets $J=\{j_0,j_1,\cdots,j_{k-1}\}$ and $I=\{i_0,i_1,\cdots,i_{m-k-1}\}$, such that $I\cap J=\emptyset$ and $I\cup J=\{0,1,\cdots,m-1\}$. Let $f: \Z_2^m\rightarrow\Z_q$ be a generalized Boolean function in $m$ variables $x_0,x_1,\cdots,x_{m-1}$. Write $\bix=(x_{j_0}\,x_{j_1}\, \cdots\, x_{j_{k-1}})$ and suppose that for each $\bid\in\Z_2^k$ the restricted functions $f|_{\bix=\bid}$ are of the form (\ref{eqn:pairs}). Define $a_\bid$ to be either $\pi(i_0)$ or $\pi(i_{m-k-1})$ in the expression $f|_{\bix=\bid}$ in (\ref{eqn:pairs}) and write
\beq 
e=\sum_{\bid\in\Z_2^k}x_{a_\bid}\prod_{\alpha=0}^{k-1}x_{j_\alpha}^{d_\alpha}(1-x_{j_\alpha})^{(1-d_\alpha)}.
\eeq
Then the polyphase vectors associated with the functions
\beq
f\,+\,\frac{q}{2}\left(\sum\limits_{\alpha=0}^{k-1}c_\alpha\,x_{j_\alpha}+c'\,e\right) \quad c_0,\cdots,c_{k-1},c'\in \Z_2
\eeq
form a complementary set of size $2^{k+1}$.
\end{theorem}
\par
\begin{proof}
Write $\bic=(c_0\,c_1\,\cdots\,c_{k-1})$ and denote the $2^{k+1}$ vectors in the complementary set as $\biF_{\bic c'}$. We have to show that the sum of auto-correlations $\sum_{\bic,\,c'} A\left(\biF_{\bic c'}\right)(\ell)$  is zero for $\ell\ne 0$.
We employ Lemma \ref{lem:CCF-Expanding} and write
\begin{align*}
&\sum_{\bic,\,c'} A(\biF_{\bic c'})(\ell)
=\sum\limits_{\bic,\,c'}\sum\limits_{\bid}A(\biF_{\bic c'}|_{\bix=\bid})(\ell)\\
&+\sum\limits_{\bic,\,c'}\sum\limits_{\bid_1\ne\bid_2}C(\biF_{\bic c'}|_{\bix=\bid_1},\biF_{\bic c'}|_{\bix=\bid_2})(\ell)=S_1+S_2.
\end{align*}
We first focus on the term $S_1$, which becomes
\beq
S_1=\sum\limits_{\bic}\sum\limits_{\bid}\left(A(\biF_{\bic 0}|_{\bix=\bid})(\ell)+A(\biF_{\bic 1}|_{\bix=\bid})(\ell)\right).
\eeq
Note that $e|_{\bix=\bid}=x_{a_\bid}$. Thus the functions corresponding to $\biF_{\bic 0}|_{\bix=\bid}$ and $\biF_{\bic 1}|_{\bix=\bid}$ are
\beq 
f|_{\bix=\bid}+\frac{q}{2}\sum\limits_{\alpha=0}^{k-1}c_\alpha\,d_\alpha
\quad\mbox{and}\quad
f|_{\bix=\bid}+\frac{q}{2}\sum\limits_{\alpha=0}^{k-1}c_\alpha\,d_\alpha+\frac{q}{2}\,x_{a_\bid},
\eeq
respectively. Notice that the sum over $\alpha$ is just a constant occuring in both functions. Hence, by hypothesis and by Theorem \ref{thm:pairs}, $\biF_{\bic 0}|_{\bix=\bid}$ and $\biF_{\bic 1}|_{\bix=\bid}$ form a complementary pair. It follows that the inner term of $S_1$ is zero for $\ell\ne 0$, and thus, also $S_1$ itself is zero for $\ell\ne 0$.
\par
Next we focus on the term $S_2$ and rearrange the sum as follows
\beq
S_2=\sum\limits_{\bid_1\ne\bid_2}\sum\limits_{c'}\sum\limits_{\bic}C(\biF_{\bic c'}|_{\bix=\bid_1},\biF_{\bic c'}|_{\bix=\bid_2})(\ell).
\eeq
For fixed $\bid_1$, $\bid_2$, and $c'$ we consider the inner sum. The functions corresponding to $\biF_{\bic c'}|_{\bix=\bid_1}$ and  $\biF_{\bic c'}|_{\bix=\bid_2}$ are
\beqn
\left(f+\frac{q}{2}\,c'\,e\right)\bigg|_{\bix=\bid_1}\!\!\!+\frac{q}{2}h_1
\;\;\mbox{and}\;\;
\left(f+\frac{q}{2}\,c'\,e\right)\bigg|_{\bix=\bid_2}\!\!\!+\frac{q}{2}h_2,
\label{eqn:inner-functions}
\eeqn
respectively, where
\beq
h_1=\sum\limits_{\alpha=0}^{k-1}c_\alpha\,d_{1,\alpha}
\quad\mbox{and}\quad
h_2=\sum\limits_{\alpha=0}^{k-1}c_\alpha\,d_{2,\alpha}.
\eeq
Let us consider the terms $h_1$ and $h_2$ themselves as Boolean functions in the variables $c_0,c_1,\cdots,c_{k-1}$. 
Since $h_1$ and $h_2$ are multiplied with $q/2$ in (\ref{eqn:inner-functions}), an inversion of $h_1$ and $h_2$ implies a sign change of $\biF_{\bic c'}|_{\bix=\bid_1}$ and $\biF_{\bic c'}|_{\bix=\bid_2}$, respectively. Now write $g_1=(f+q/2\,c'\,e)|_{\bix=\bid_1}$ and $g_2=(f+q/2\,c'\,e)|_{\bix=\bid_2}$, and let $\biG_1$ and $\biG_2$ be their associated vectors, respectively. Then the inner sum of $S_2$ comprises terms of the form $C(\pm\, \biG_1,\pm\, \biG_2)(\ell)$, where $C(+\biG_1,+\biG_2)(\ell)$ and $C(-\biG_1,-\biG_2)(\ell)$ occur if $h_1=h_2$, and $C(+\biG_1,-\biG_2)(\ell)$ and $C(-\biG_1,+\biG_2)(\ell)$ occur if $h_1\ne h_2$. It is easy to show that  $C(+\biG_1,+\biG_2)(\ell)=$ $C(-\biG_1,-\biG_2)(\ell)=$ $-C(+\biG_1,-\biG_2)(\ell)=$ $-C(-\biG_1,+\biG_2)(\ell)$. In order to prove that the inner sum of $S_2$ is zero, we have to show that $h_1=h_2$ and $h_1\ne h_2$ occur equally often as $\bic$ runs through all possible values. Recall that $\bid_1\ne \bid_2$. Hence the difference 
\beq
h_2-h_1=\sum\limits_{\alpha=0}^{k-1}c_\alpha\,(d_{2,\alpha}-d_{1,\alpha})
\eeq
is a nonzero linear Boolean function in the variables $c_0,c_1,\cdots,c_{k-1}$. According to the randomization lemma \cite[page 372]{MacWilliams1977}, such a function produces the values '0' and '1' equally often as $\bic$ takes all possible values. Thus $h_1$ and $h_2$ are distinct for half of all cases. It follows that the inner sum of $S_2$, and hence, also $S_2$ itself is zero for all $\ell$.
\end{proof}
\par
The following corollary is a direct consequence of Theorem \ref{thm:sets} and Result \ref{res:sets} and provides a general upper bound on the PMEPR of polyphase sequences of length $2^m$.
\begin{corollary}
\label{cor:PMEPR}
Let $f\,:\,\Z_2^m\rightarrow\Z_q$ be a generalized Boolean function. If there exists a set of $k$ variables $\bix=(x_{j_0}\,x_{j_1}\cdots x_{j_{k-1}})$, such that for each $\bid\in\Z_2^k$ the function $f|_{\bix=\bid}$ is of the form (\ref{eqn:pairs}), then the PMEPR of the polyphase vector associated with $f$ is at most $2^{k+1}$.
\end{corollary}


\section{Reed--Muller Codes and Generalizations}
\label{sec:RM-codes}
A code $\C$ of length $n$ over the ring $\Z_q$ is defined as a subset $\C\subseteq$ $\Z_q^n$. Such a code $\C$ is said to be linear if each $\Z_q$-linear combination of the codewords of $\C$ yields again a codeword of $\C$. Let $\bia$ be a codeword of $\C$. The Hamming weight $\wt_H(\bia)$ is defined as the number of nonzero entries in $\bia$. The Lee weight of $\bia$ is defined as $\wt_L(\bia)=\sum_{i=0}^{n-1}\min(a_i,q-a_i)$. For linear codes the minimum Hamming distance $\dist_H(\C)$ (minimum Lee distance $\dist_L(\C)$) of a code $\C$ is defined as the minimum Hamming (Lee) weight of the nonzero codewords of $\C$. We next consider codes defined by generalized Boolean functions. 
\begin{definition}
The Reed--Muller code $\RM(r,m)$ of order $r$ and length $2^m$ is the set of all binary vectors that can be associated with a Boolean function of order at most $r$.
\end{definition}
\par
The code $\RM(r,m)$ is linear, comprises $2^{\sum_{i=0}^r{m\choose i}}$ codewords, and has minimum Hamming (and Lee) distance $2^{m-r}$. For further details see \cite{MacWilliams1977}. Next we define a new generalization of the classical Reed--Muller codes. Notice that in the following we restrict $q$ to be a power of $2$, i.e. $q=2^h$.
\begin{definition}
\label{def:ZRM}
For $h>p$ and $r\ge p$ we define the code $\ZRM_{2^h}^p(r,m)$ as the set of all vectors of length $2^m$ that can be associated with a generalized Boolean function $\Z_2^m\rightarrow\Z_{2^h}$ comprising the monomials of order at most $r-p$ and $2^i$ times the monomials of order $r-p+i$ with $i=1,2,\dots,p$.
\end{definition}
\par
Apparently the code $\ZRM_{2^h}^p(r,m)$ is linear, and a simple counting argument shows that
\beqn
\log_2\big|\ZRM_{2^h}^p(r,m)\big|=\sum_{i=0}^{r-p}h{m\choose i}+\sum_{i=1}^{p}(h-i){m\choose i+r-p}.
\label{eqn:dim-ZRM}
\eeqn
We remark that the code $\ZRM_{2^h}^p(r,m)$ generalizes the codes $\RM_{2^h}(r,m)$ and $\ZRM_{2^h}(r,m)$ from \cite{Davis1999}. The code $\RM_{2^h}(r,m)$ simply reads $\ZRM_{2^h}^0(r,m)$ and $\ZRM_{2^h}(r,m)$ is in our notation $\ZRM_{2^h}^1(r,m)$. For $p>1$ the code $\ZRM_{2^h}^p(r,m)$ yields a new generalization of the Reed--Muller code, that was, to our best knowledge, not mentioned before.
\par
\begin{theorem}
\label{thm:dHL}
The minimum Hamming distance of $\ZRM_{2^h}^p(r,m)$ is equal to $2^{m-r}$ and the minimum Lee distance of $\ZRM_{2^h}^p(r,m)$ is equal to $2^{m-r+p}$.
\end{theorem}
\begin{proof}
Since $\ZRM_{2^h}^p(r,m)$ is linear, we need to find the minimum weights of the nonzero codewords. We first prove a lower bound for the weights. Then we show that there exists at least one codeword that attains this bound. 
\par
The proof is by induction on $p$ and $h$, where we take the statement in the above theorem as a hypothesis. The base case for the induction is $p=0$ and $h=1$. Then $\ZRM_2^0(r,m)$ is equal to $\RM(r,m)$ and has minimum Hamming and Lee weight $2^{m-r}$. Suppose $\bia=(a_0\,a_1\dots a_{n-1})$ is a codeword of $\ZRM_{2^h}^p(r,m)$ and let $\bib=(b_0\,b_1\dots b_{n-1})$ with $b_i=a_i \bmod 2^{h-1}$ be codeword over $\Z_{2^{h-1}}$. We will use the easily verified inequalities $\wt_H(\bia)\ge\wt_H(\bib)$ and $\wt_L(\bia)\ge\wt_L(\bib)$. The first relation is immediately clear and the latter one follows because $a_i\in\{b_i,b_i+2^{h-1}\}$ and $\min(a_i,2^h-a_i)\ge\min(b_i,2^{h-1}-b_i)$.
\par {\itshape Case 1:}
$\bib=\bnull$. In this case, $\bia$ comprises only values of either $0$ or $2^{h-1}$. Then $2^{1-h}\,\bia$ is a codeword of $\RM(r,m)$. Hence $\wt_H(\bia)\ge2^{m-r}$ and $\wt_L(\bia)\ge 2^{h-1}\,2^{m-r}\ge 2^{m-r+p}$, since by Definition \ref{def:ZRM}, $h>p$.
\par {\itshape Case 2:}
$\bib\ne \bnull$ and $h=p+1$. Now $\bib$ is a nonzero codeword of $\ZRM_{2^{h-1}}^{p-1}(r-1,m)$. Let us first consider the case $p=1$. Then we have $h=2$. Hence $\bib$ belongs to $\ZRM_2^0(r-1,m)$, which is equal to $\RM(r-1,m)$. Thus we have $\wt_H(\bia)\ge\wt_H(\bib)\ge2^{m-r+1}$ and $\wt_L(\bia)\ge\wt_L(\bib)\ge2^{m-r+1}$.
Now consider $p>1$. Then, by induction, we obtain $\wt_H(\bia)\ge\wt_H(\bib)\ge2^{m-r+1}$ and $\wt_L(\bia)\ge\wt_L(\bib)\ge2^{m-r+p}$.
\par {\itshape Case 3:}
$\bib\ne \bnull$ and $h>p+1$. In this case $\bib$ is a nonzero codeword of $\ZRM_{2^{h-1}}^p(r,m)$. By induction we eventually arrive at Case 1 or 2, and thus, we have $\wt_H(\bib)\ge2^{m-r}$ and $\wt_L(\bib)\ge2^{m-r+p}$. 
This implies that $\wt_H(\bia)\ge2^{m-r}$ and $\wt_L(\bia)\ge2^{m-r+p}$.
\par
Now consider the codeword corresponding to the Boolean function $2^p\,x_0x_1\cdots x_{r-1}$. This codeword has Hamming weight $2^{m-r}$ and Lee weight $2^{m-r+p}$. These weights attain the lower bounds derived above, which completes the proof.
\end{proof}


\section{OFDM Codes with Low PMEPR}
\label{sec:ofdm-codes}
We define two fixed lists of indices $I=\{i_0\,i_1\,\cdots\,i_{m-k-1}\}$ and $ J=\{j_0\,j_1\,\cdots\,j_{k-1}\}$ such that $I\,\cap\, J=\emptyset$ and $I\,\cup\, J=\{0,1,\cdots,m-1\}$. Suppose $g_0,g_1,\cdots,g_{m-k-1},g': \Z_2^k\rightarrow\Z_q$ are $m-k+1$ generalized Boolean functions in $k$ variables. Let $a: \Z_2^m\rightarrow\Z_{2^h}$, a generalized Boolean function in $m$ variables, be given by
\beqn
a=\sum_{\alpha=0}^{m-k-1}x_{i_\alpha}g_\alpha(x_{j_0},\cdots,x_{j_{k-1}})+g'(x_{j_0},\cdots,x_{j_{k-1}}).
\label{eqn:def-a}
\eeqn
Clearly the set of vectors that can be associated with a function of type (\ref{eqn:def-a}) forms a linear subcode of $\ZRM_{2^h}^0(k+1,m)$. Denote this subcode as $\L_{2^h}(k,m)$. Notice that $\L_{2^h}(0,m)$ is identical to $\ZRM_{2^h}^0(1,m)$, the first-order generalized Reed--Muller code.
\par
Now suppose $d$ has binary expansion $(d_0\,d_1\cdots d_{k-1})$ and let $\pi_0,\pi_1,\cdots,\pi_{2^k-1}$ be $2^k$ permutations of $\{i_0,i_1,\cdots,i_{m-k-1}\}$. Then consider the functions
\beqn
b=2^{h-1}\sum_{d=0}^{2^k-1}\sum_{\alpha=0}^{m-k-2}x_{\pi_d(\alpha)}x_{\pi_d(\alpha+1)}\prod_{\beta=0}^{k-1}x_{j_\beta}^{d_\beta}(1-x_{j_\beta})^{1-d_\beta},
\label{eqn:def-b}
\eeqn
and let the set of all vectors corresponding to a form of type (\ref{eqn:def-b}) form a code $\R_{2^h}(k,m)$. Clearly the order of $b$ is at most $k+2$, and the elements of the associated $\Z_{2^h}$-valued codewords are either $0$ or $2^{h-1}$. Since a codeword is completely determined by the $2^k$ permutations $\pi_0,\pi_1,\cdots,\pi_{2^k-1}$, there exist $[(m-k)!/2]^{2^k}$ codewords in $\R_{2^h}(k,m)$.
\par
Now suppose $\bib$ is a codeword in $\R_{2^h}(k,m)$. Then the set of codewords
\beqn
\{\bib+\bia\,|\,\bia\in\L_{2^h}(k,m)\}
\label{eqn:coset}
\eeqn
is a coset of the linear code $\L_{2^h}(k,m)$ with $\bib$ being its coset representative.
\begin{theorem}
Let $\bib\in\R_{2^h}(k,m)$. Then each polyphase codeword of the coset (\ref{eqn:coset}) has PMEPR at most $2^{k+1}$.
\end{theorem}
\begin{proof}
Consider the notations above. Let $a$ and $b$ be the generalized Boolean functions corresponding to the codewords $\bia$ and $\bib$, respectively. Recall the definition of $I$ and $ J$ and write $\bix=(x_{j_0}\,x_{j_1}\cdots x_{j_{k-1}})$. According to Corollary \ref{cor:PMEPR} we have to show that for each binary vector $\bid=(d_0\,d_1\cdots d_{k-1})$ the expression $(a+b)|_{\bix=\bid}=a|_{\bix=\bid}+b|_{\bix=\bid}$ is of the form (\ref{eqn:pairs}). 
Considering (\ref{eqn:def-a}) we have
\beq
a|_{\bix=\bid}=\sum_{\alpha=0}^{m-k-1}x_{i_\alpha}g_\alpha|_{\bix=\bid}+g'|_{\bix=\bid},
\eeq
where $g_0,g_1,\dots,g_{k-1}$, and $g'$ are arbitrary generalized Boolean functions in the variables $x_{j_0},x_{j_1},\cdots,x_{j_{k-1}}$. Apparently, after restriction in $\bix$, each of the restricted functions $g_\alpha|_{\bix=\bid}$ and $g'|_{\bix=\bid}$ are constants, and $a|_{\bix=\bid}$ becomes an affine function. Now consider $b|_{\bix=\bid}$. It is easy to verify that
\beq
b|_{\bix=\bid}=2^{h-1}\sum_{\alpha=0}^{m-k-2}x_{\pi_d(\alpha)}x_{\pi_d(\alpha+1)},
\eeq
which has the form of the quadratic part in (\ref{eqn:pairs}). Thus, by Corollary \ref{cor:PMEPR}, the polyphase vectors associated with $(a+b)$ have PMEPR at most $2^{k+1}$.
\end{proof}
\par
Before we state our new constructions let us define subcodes of $\L_{2^h}(k,m)$. Let $\A_{2^h}^p(k,r,m)=\L_{2^h}(k,m)\cap\ZRM_{2^h}^p(r,m)$. Of course $\A_{2^h}^p(k,r,m)$ is linear and  $\A_{2^h}^0(k,k+1,m)=\L_{2^h}(k,m)$. By inspecting (\ref{eqn:def-a}) and using (\ref{eqn:dim-ZRM}), we have
\begin{align}
\label{eqn:enc_bits}
&\log_2\!\big|\A_{2^h}^p(k,r,m)\big|\!=\!\sum_{i=0}^{r-p}h{k\choose i}\!+\!\sum_{i=1}^p(h-i){k\choose i+r-p}\nonumber\\
&+(m-k)\left(\sum_{i=0}^{r-p-1}h{k\choose i}+\sum_{i=1}^{p}(h-i){k\choose i+r-p-1}\right),
\end{align}
Now we use cosets of the code $\A_{2^h}^p(k,r,m)$ to construct three code classes with PMEPR at most $2^{k+1}$.
\par
{\itshape Class I Codes:} 
A very simple code can be constructed by using just a single coset of $\A_{2^h}^p(k,r,m)$, i.e.\ 
\beq
\{\bib+\bia\,|\,\bia\in\A_{2^h}^p(k,r,m)\},\qquad \bib\in\R_{2^h}(k,m).
\eeq
Clearly the number of encodable bits is given by (\ref{eqn:enc_bits}). 
Since the constant offset leaves the distance properties unchanged and $\A_{2^h}^p(k,r,m)\subseteq\ZRM_{2^h}^p(r,m)$, this code has minimum Hamming distance $2^{m-r}$ and minimum Lee distance $2^{m-r+p}$.
\par
{\itshape Class II Codes:} 
Consider the functions corresponding to $\R_{2^h}(k,m)$, and set $\pi=\pi_0\!=\!\cdots\!=\!\pi_{2^k-1}$ in the definition of those functions in (\ref{eqn:def-b}). Then we obtain quadratic forms of type
\beq
b'=2^{h-1}\sum_{\alpha=0}^{m-k-2}x_{\pi(\alpha)}x_{\pi(\alpha+1)}.
\eeq
There exist $(m-k)!/2$ vectors associated with such a quadratic form. Let $\R'_{2^h}(k,m)$ denote this set. Then we define the code
\beq
\bigcup_{\bib\in\R'_{2^h}(k,m)}\{\bib+\bia\,|\,\bia\in\A_{2^h}^p(k,r,m)\}.
\eeq
For $k>0$ and $r>1$ the above code is a union of $(m-k)!/2$ cosets of $\A_{2^h}^p(k,r,m)$ inside $\ZRM_{2^h}^p(r,m)$, and hence, it has minimum Hamming distance $2^{m-r}$ and minimum Lee distance $2^{m-r+p}$. For $k=0$ and $r=1$ the code is a union of $m!/2$ cosets of $\ZRM_{2^h}^p(1,m)$ inside $\ZRM_{2^h}^{p+1}(2,m)$. In this case it has minimum Hamming distance $2^{m-2}$ and minimum Lee distance $2^{m-1+p}$. The maximal number of encodable bits amounts to $\log_2\big|\A_{2^h}^p(k,r,m)\big|+\lfloor\log_2(m-k)!/2\rfloor$.
\par
{\itshape Class III Codes:} 
Recall that (\ref{eqn:def-b}) identified $((m-k)!/2)^{2^k}$ coset represenatives. Note that these coset representatives have order at most $k+2$ and its elements are either $0$ or $2^{h-1}$. Then, for $p>0$, we define the Class III codes as follows
\beq
\bigcup_{\bib\in\R_{2^h}(k,m)}\{\bib+\bia\,|\,\bia\in\A_{2^h}^{p-1}(k,k+1,m)\}.
\eeq
This code is a union of $((m-k)!/2)^{2^k}$ cosets of $\A_{2^h}^{p-1}(k,k+1,m)$ inside $\ZRM_{2^h}^p(k+2,m)$. Hence it has minimum Hamming distance $2^{m-k-2}$ and minimum Lee distance $2^{m-k-2+p}$. With such a code one can encode $\log_2\big|\A_{2^h}^{p-1}(k,k+1,m)\big|+\lfloor2^k\log_2 (m-k)!/2\rfloor$ bits.
\par
\enlargethispage{-3ex}
{\itshape Remarks:} Some relations to previous constructions are given below.
\vspace{-1ex}
\begin{trivlist}
\item[ 1)] Setting $k=0$ and $p=0$ in the Class II codes results in codes that coincide with those constructed in \cite{Davis1999}. Then the codes comprise $m!/2$ cosets of $\ZRM_{2^h}^0(1,m)$ inside $\ZRM_{2^h}^1(2,m)$ and have PMEPR at most $2$. 
\item[ 2)] Setting $p=1$ in the Class III codes, then each codeword can be obtained by interleaving $2^k$ codewords of length $m-k$ from the codes considered in \cite{Davis1999} (see 1)).
\item[ 3)] If $\A^0_{2^h}(k,2,m)$ is chosen as the underlying code for the Class II codes, the resulting codes are similar to those considered in \cite{Paterson2000a}. Then we obtain a subcode of $\ZRM_{2^h}^0(2,m)$. The difference between our construction and that in \cite{Paterson2000a} is that we apply the permutation of the variable indices only to those indices from the set $I$. This way we can guarantee that the codewords are generated exactly once, since all quadratic forms corresponding to the coset represenatives are permutation invariant. In contrast to that, in \cite{Paterson2000a} the permutation was applied to the indices $\{0,1,\dots,m-1\}$. Then, in order to avoid multiple generations of codewords, it was necessary to introduce some constraints on the quadratic parts of the associated functions, which made the handling of those codes a bit unwieldy.
\end{trivlist}
\vspace{-1ex}
\par
Notice also that, contrarily to the codes considered in \cite{Davis1999} and \cite{Paterson2000a}, our codes are not unions of cosets of a generalized first-order Reed--Muller code $\ZRM_{2^h}^0(1,m)$, but instead are unions of cosets of the linear code $\A_{2^h}^p(k,r,m)$, which has, for $k>0$ and $p>0$, more codewords than $\ZRM_{2^h}^0(1,m)$. Hence, compared to the approaches in \cite{Davis1999} and \cite{Paterson2000a}, we need less cosets to achieve about the same code rate. This way the encoding and decoding procedures become simpler, in particular for larger block lengths. For the details about encoding and decoding of the proposed code classes we refer the reader to \cite{Schmidt2005c}.


\section{Conclusion}
\label{sec:conclusion}
In this paper a large family of sequences lying in complementary sets have been presented. Moreover the classical Reed--Muller codes have been generalized in a novel manner. We have shown that the family of sequences lying in complementary sets form cosets of a linear code, which are contained in the generalized Reed--Muller code $\ZRM_{2^h}^p(r,m)$. This way new codes for OFDM with low PMEPR have been proposed, which are not limited to be a subcode of the second-order generalized Reed--Muller code. A number of code families has been presented, where PMEPR, code rate, minimum distance, and encoding/decoding complexity can be traded against each other.


\bibliographystyle{ieeetr}
\bibliography{references}

\begin{thebibliography}{1}

\bibitem{Jones1996}
A.~E. Jones and T.~A. Wilkinson, ``Combined coding for error control and
  increased robustness to system nonlinearities in {OFDM},'' {\em Proc. of IEEE
  46th Vehicular Technology Conference (VTC)}, Apr. 1996.

\bibitem{Davis1999}
J.~A. Davis and J.~Jedwab, ``Peak-to-mean power control in {OFDM}, {Golay}
  complementary sequences, and {Reed-Muller} codes,'' {\em IEEE Trans. Inform.
  Theory}, vol.~45, pp.~2397--2417, Nov. 1999.

\bibitem{Golay1961}
M.~J.~E. Golay, ``Complementary series,'' {\em IRE Trans. Inform. Theory},
  1961.

\bibitem{Paterson2000a}
K.~G. Paterson, ``Generalized {Reed-Muller} codes and power control in {OFDM}
  modulation,'' {\em IEEE Trans. Inform. Theory}, vol.~46, pp.~104--120, Jan.
  2000.

\bibitem{Tseng1972}
C.-C. Tseng and C.~L. Liu, ``Complementary sets of sequences,'' {\em IEEE
  Trans. Inform. Theory}, vol.~18, pp.~644--652, Sep. 1972.

\bibitem{Schmidt2005}
K.-U. Schmidt and A.~Finger, ``Constructions of complementary sequences for
  power-controlled {OFDM} transmission,'' {\em Proc. of Workshop on Coding and
  Cryptography (WCC), Bergen, Norway}, 2005.

\bibitem{MacWilliams1977}
F.~J. MacWilliams and N.~J.~A. Sloane, {\em The Theory of Error-Correcting
  Codes}.
\newblock North Holland Mathematical Library, 1977.

\bibitem{Stinchcombe2000}
T.~E. Stinchcombe, {\em Aperiodic Autocorrelations of Length {$2^m$} Sequences,
  Complementarity, and Power Control for {OFDM}}.
\newblock PhD thesis, University of London, Apr. 2000.

\bibitem{Schmidt2005c}
K.-U. Schmidt, ``Complementary sets, generalized {Reed--Muller} codes, and
  power control for {OFDM},'' {\em submitted}, 2005.

\end{thebibliography}


\end{document}